\documentclass[aps,amsfonts,pra,twocolumn,showpacs]{revtex4}

\usepackage{subfigure}
\usepackage{color}
\usepackage{textcomp}
\usepackage{graphicx}
\usepackage{amssymb}
\usepackage{amsmath}
\usepackage{bm}
\usepackage{amsmath, amsthm, amssymb}
\usepackage{multirow}

\newcommand{\dhch}[1]{{\color{black} #1}}
\newcommand{\sgch}[1]{{\color{black} #1}}

\def\beq{\begin{equation}}
\def\eeq{\end{equation}}
\def\bea{\begin{eqnarray}}
\def\eea{\end{eqnarray}}

\begin{document}

\title{Griffiths effects and slow dynamics in nearly many-body localized systems}

\author{Sarang Gopalakrishnan}
\affiliation{Department of Physics and Walter Burke Institute, California Institute of Technology, Pasadena, CA, USA}

\author{Kartiek Agarwal}
\affiliation{Department of Physics, Harvard University, Cambridge, MA, USA}

\author{Eugene A. Demler}
\affiliation{Department of Physics, Harvard University, Cambridge, MA, USA}

\author{David A. Huse}

\affiliation{Princeton University, and Institute for Advanced Study, Princeton, NJ, USA}

\author{Michael Knap}

\affiliation{Department of Physics, Walter Schottky Institute, and Institute for Advanced Study, Technical University of Munich, 85748 Garching, Germany}


\begin{abstract}
The low-frequency response of systems near a many-body localization transition can be dominated by rare regions that are locally critical or ``in the other phase''.
It is known that, in one dimension, these rare regions can cause the d.c. conductivity and diffusion constant to vanish even inside the delocalized thermal phase.
Here, we present a general analysis of such Griffiths effects in the thermal phase near the many-body localization transition: we consider both one-dimensional and
higher-dimensional systems, subject to 
quenched randomness, 
and discuss both linear response (including the frequency- and wavevector-dependent conductivity) and more general dynamics.
In all the regimes we consider, we identify observables that are dominated by rare-region effects.  In some cases (one-dimensional systems and Floquet systems
with no extensive conserved quantities), 
essentially all long-time local observables are dominated by rare-region effects; in others, generic observables are instead dominated by hydrodynamic long-time tails throughout the thermal phase, and one must look at \emph{specific} probes, such as spin echo, to see Griffiths behavior.

\end{abstract}

\maketitle

\section{Introduction}

The many-body localization (MBL) transition is a phase transition, occurring in isolated and usually disordered interacting quantum many-body systems, at which equilibrium statistical
mechanics breaks down \cite{pwa,baa,oh,znidaric,ph,nh}.  On one side of the transition (in the ``thermal phase'')
the system comes to thermal equilibrium under its own unitary dynamics;
on the other side (in the ``MBL phase''), it does not, acting instead as a ``quantum memory''~\cite{spa, hno, bvav,ckls,hnops,bn,rms}. A considerable amount
of numerical and experimental evidence supports the existence of these two distinct phases~\cite{oh, ph,kbp,iyer,kondov,shahar2014, bloch, smith}; in addition,
the existence of the MBL phase in certain one-dimensional systems can be proven with minimal assumptions \cite{jzi}.
Although some properties of both the MBL and
thermal phases away from the transition are believed to be phenomenologically understood, these phenomenological approaches
(the ``l-bit'' model for the MBL phase~\cite{spa,hno, spa2}, and equilibrium transport theory and hydrodynamics for the thermal phase)
are mutually incompatible, and both break down as the transition is approached.  Hence many basic open questions remain about the
behavior near and at the MBL phase transition.
%

The numerical evidence, from the exact diagonalization of small systems, suggests that the MBL transition in one dimension in systems
with quenched randomness is governed by an infinite-randomness critical point \cite{ph}, and that the regimes near the transition are ``Griffiths''
regimes, in the sense that their low-frequency response is dominated by the contributions from rare regions \cite{agkmd,VHA,ppv,mbmott}.
In particular, the thermal phase near the transition exhibits anomalous \dhch{(sub)}diffusion \cite{blcr, agkmd}, as well as anomalous spectral correlations~\cite{santos, SM}, whereas the low-frequency
conductivity just on the localized side of the transition goes as $\sigma(\omega) \sim \omega$~\cite{mbmott}.
These features are naturally explained in terms of the following physical picture: a system near the MBL transition is highly inhomogeneous,
and can be regarded as a patchwork of locally thermalizing
 and locally insulating regions. When the system is globally in the thermal phase, its transport is (in one dimension) blockaded by rare insulating segments,
 giving rise to anomalous diffusion. By contrast, when the system is globally in the insulating phase, 
 its low-frequency response is dominated by locally thermalizing \dhch{(or critical)}
 islands and their surroundings.

\begin{table}[b]
\centering
\caption{Summary of main qualitative results, indicating regimes in which Griffiths effects are dominant and subleading.}
\label{my-label}
\begin{tabular}{|l|l|l|l|}
\hline
{\it Griffiths effects in...}                                                                      &          & {\bf Hamiltonian} & {\bf Floquet} \\ \hline
\multirow{2}{*}{\begin{tabular}[c]{@{}l@{}}Generic spatially\\ averaged response\end{tabular}}     & 1D       & Leading           & Leading       \\ \cline{2-4}
                                                                                                   & Higher D & Subleading        & Leading       \\ \hline
Averaged spin echo                                                                                 & Any D    & Leading           & Leading       \\ \hline
\multirow{2}{*}{\begin{tabular}[c]{@{}l@{}}Typical response\\ (generic or spin echo)\end{tabular}} & 1D       & Leading           & Leading       \\ \cline{2-4}
                                                                                                   & Higher D & Subleading        & Subleading    \\ \hline
\end{tabular}
\end{table}

The existing work on Griffiths effects near the MBL transition has focused primarily on transport in systems with quenched disorder (although the dynamics of contrast decay is briefly discussed in Ref.~\cite{ppv}, whose conclusions agree with ours). Moreover, the discussion of the thermal side has been restricted to one dimension. However, ongoing experiments with ultracold atomic systems~\cite{bloch} are \emph{not} limited to one dimension, and are most naturally probed through quench dynamics and interferometry rather than transport.  It is the objective of this paper to explore Griffiths effects in these more general settings: to extend previous results from dimension $d=1$ to $d>1$ and from transport to more general dynamics.  We only consider states that correspond to nonzero (and sometimes infinite) temperature. Also, when we consider $d>1$, we are making the assumption that the MBL phase can exist as a truly distinct dynamical quantum phase in $d>1$, although the existing proof \cite{jzi} of the existence of MBL is limited to the case of $d=1$. Regardless of whether strict MBL exists in $d > 1$, however, our results should apply at intermediate times.

Our main focus in this paper is on the autocorrelation functions of 
local operators: these can be related to transport, but also to noise~\cite{adm}, interferometric measurements~\cite{kkg}, and quench dynamics (as discussed below).
We find that, in general, for the spatially-averaged equilibrium autocorrelation function of most local operators $O$,
rare critical or insulating regions in the thermal phase give a contribution to the long-time behavior of the form:
\beq\label{eq1}
\langle O(t) O(0) \rangle - \langle O \rangle^2 \sim \exp(-\alpha \log^d t) ~,
\eeq
where $\alpha$ is a nonuniversal, observable-dependent constant that varies continuously through the thermal phase and goes
to zero at the MBL transition.  This result applies for any operator $O$ that ``freezes'' in the MBL phase, in the sense that its autocorrelation does
not decay to zero in the MBL phase.
%
The behavior~\eqref{eq1} is power-law in one dimension, but faster than a power-law in higher dimensions.  Thus, in higher dimensions, Griffiths effects are generically subleading to hydrodynamic power-laws; however, we identify specific observables (such as spin echo) as well as systems (``fully generic'' Floquet systems with no conserved densities) for which hydrodynamic power laws are absent and Griffiths effects are therefore dominant.
In addition to the rare-region contribution to \emph{averaged} autocorrelation functions, in one dimension they can dominate autocorrelation functions at a \emph{typical} point~\cite{fntyp}, by acting as bottlenecks as discussed in Refs.~\cite{agkmd, VHA}.  In higher dimensions, this effect is absent.  These various regimes are summarized in Table I.

Many-body localization can also occur in systems without quenched randomness that are subject to quasiperiodic potentials \cite{iyer,bloch}.  Within the MBL phase, both quasiperiodic and random systems can be subject to a different type of Griffiths effects due to rare regions of the {\it state} that locally take the state to a many-body mobility edge~\cite{mbmott}, if such a mobility edge is present (as suggested in Refs.~\cite{baa, kbp, lfa2015, bshb}, but see also Ref.~\cite{mobilityedge}).  Since the MBL phase is frozen, such rare regions of the state are dynamically stable and thus behave like quenched randomness.  But in the thermal phase this cannot happen: a rare region of the state that takes it locally in to the insulating phase will not be stable, but instead will be ``melted'' (thermalized) by the surrounding thermal environment.  Thus we do not expect dynamic Griffiths effects in the thermal phase of nonrandom quasiperiodic systems, where there are no rare regions of the Hamiltonian (or Floquet operator).

This work is arranged as follows. In Sec.~\ref{notation}, we introduce our notation and assumptions. In Sec.~\ref{review} we summarize previous results on
one-dimensional Griffiths effects. In Sec.~\ref{floquet}, we discuss Griffiths effects in the conceptually simplest case: that of a Floquet system that has no
extensive conservation laws.  In Sec.~\ref{Hamiltonian} we turn to systems with global conservation laws in general dimensions, and discuss the competition between hydrodynamic long-time tails and Griffiths effects. We find that, for generic autocorrelation functions, the Griffiths effects are subleading in dimensions greater than one, and identify specific observables---in particular, the spin echo response (Sec.~\ref{secse})---that remain dominated by Griffiths effects in all dimensions.
In Sec.~\ref{dominantregions} we consider the nature of the dominant rare regions; this discussion addresses the behavior of the prefactor $\alpha$ in Eq.~\eqref{eq1} near the transition. Finally, Sec.~\ref{summary} summarizes our results.

\section{Notation and assumptions}\label{notation}

We first set out some general assumptions and introduce some notation that we shall use throughout the paper.
%
We consider systems that have one, or a few, extensive conserved scalar quantities (e.g., energy, charge, and/or spin-projection along some axis),
but no other special symmetries, as well as
fully generic Floquet systems, in which there are no extensive conserved quantities.
We take the interactions to be short-range in space.
We take the disorder to be spatially uncorrelated (or to have a correlation length that is short compared with the length scales of interest to us). We assume that the system is defined on a lattice with finite on-site Hilbert space.

\sgch{We shall be primarily interested in the behavior of autocorrelation functions of generic Hermitian operators, $C(x, t) \equiv \langle O(x, t) O(x, 0) \rangle_x - \langle O(x) \rangle^2$, where $O(x)$ is an operator with finite support centered at the point $x$. The brackets $\langle \ldots \rangle$ denote averages with respect to a (presumably thermal) density matrix.
For systems that have conserved quantities, we shall also explore the autocorrelation functions of operators that are ``special,'' such as the conserved densities and their currents (denoted $j$). (For the associated autocorrelation functions we use the standard notation, such as $\sigma \sim \langle j j \rangle$ for the conductivity.) We shall address both spatially averaged and typical behavior. We denote the spatial average of $C(x, t)$ as $[C(x,t)]$, and it is defined in the obvious way. The \emph{typical} value of $C(x,t)$ is formally defined as $C_{typ}(t) \equiv \exp\{ [\log C(x,t) ] \}$. The typical and average values differ because averaging the logarithms reduces the weight of the contribution for rare regions. We shall use this formal sense of ``typical'' and its colloquial sense interchangeably: for the Griffiths effects discussed here, it is straightforward to check that these senses are indeed equivalent (i.e., rare regions do not dominate the logarithmic average).}

We denote the characteristic microscopic energy scale of the system by $W$.  The global control parameter driving the MBL transition is
denoted by $\delta(\{ \Gamma \})$, where $\Gamma$ denotes the physical parameters (energy density, interaction strength, etc.) that
affect the transition; we define $\delta$ so that $\delta=0$ at the critical point, $\delta >0$ in the thermal phase, and $\delta <0$
in the MBL phase.  We will denote the local value of $\delta$ by $\hat\delta$.
We shall assume that the MBL transition is continuous; this assumption is consistent with existing numerical evidence, but the evidence itself is mostly restricted to one dimension.

We focus on the response at times that are long (or frequencies that are small) compared with the characteristic microscopic scales of the system.
The rare regions we shall consider are correspondingly large compared with the lattice spacing, so that coarse-grained notions of the ``local properties'' are meaningful for each region.  In most of this paper, we consider rare regions whose linear size is large compared to the correlation length, which is denoted $\xi$.
Because these regions are large, one can argue on ``large deviations'' grounds~\cite{dembo} that
the probability of having some rare local property $\gamma$ behaves as $\sim \exp{(-r(\delta,\gamma)V)}$, where $V$ is the volume of the rare region,
and $r(\delta,\gamma)$ is a (non-negative) ``rate function'' that vanishes as $\gamma$ approaches $\gamma_{typ}(\delta)$, the typical behavior of a region
for the control parameter $\delta$.  For instance, if the distribution obeys the central limit theorem we expect that $r(\gamma, \delta) \sim \varphi(\delta) (\gamma - \gamma_{typ}(\delta))^2$ for small $|\gamma - \gamma_{typ}|$.  It is conceivable that the prefactor $\varphi(\delta)$ itself vanishes or diverges at the critical point, because the cost of a region with anomalously thermal or localized properties might scale nonexponentially at the critical point. If $\varphi(\delta) \sim |\delta|^\rho$ near the critical point, our conclusions are robust so long as $\rho > -1$---this includes the cases where (a) rare regions are anomalously common at the critical point, (b) the rate function is nonsingular at the critical point, and (c) rare regions are anomalously suppressed at the critical point, but the suppression is not too severe. We cannot rule out the possibility that $\rho < -1$, in which case rare regions are completely suppressed at the critical point, but as this scenario seems highly implausible we shall not consider it further.
Note that we are assuming that to make an insulating rare region, a nonzero fraction of that region has to be atypical, thus the factor of $V$ in the exponent in the probability.  This seems reasonable for rare insulating regions in the thermal phase, although for the opposite case, namely rare thermalizing regions in the MBL phase, it is less obvious that the atypical regions need to be a nonzero fraction of the total volume in the limit of such rare thermalizing regions of large volume~\cite{zzdh}.

The correlation length $\xi \sim |\delta|^{-\nu}$ as the transition is approached.  On length scales longer than $\xi$ the system's behavior is typically thermal (or MBL for $\delta <0$), while on shorter scales it is typically critical.

\section{Review of one-dimensional transport}\label{review}

We first briefly summarize previous results on Griffiths effects in the thermal phase near the MBL transition (those in the localized phase are
discussed in Ref.~\cite{mbmott}, and will not concern us here). The effect of rare ``bottlenecks'' on the spread of entanglement
in one-dimensional systems can be understood fairly simply \cite{agkmd,VHA}.  The bottlenecks are rare insulating (or, potentially, critical) regions of length $L$. The transit time across a rare insulating region increases exponentially with its length; we denote it by $t(L) \sim \exp(L/\eta)$, where $\eta$ is a quantity that decreases as the region becomes more insulating. The inclusions that serve as bottlenecks at (large) time scale $t$ are those with $L \geq \eta \log{t}$. The probability of such a bottleneck is thus $\sim\exp{(-r(\delta,\eta)L)} \sim \exp{(-\eta r(\delta,\eta)\log{t})}$, i.e., it goes as a power-law of $t$.

To find the exponent, we must optimize over all possible internal parameters for the bottlenecks: in general, locally more insulating regions
will act as more effective bottlenecks, but will also be rarer. Thus, we must optimize the quantity $\eta r(\delta, \eta)$. In one dimension,
it is believed (on numerical~\cite{ph} and renormalization-group~\cite{VHA} grounds) that $\eta$ approaches a finite value $\eta_c$ at the
critical point. Given this assumption, one can check that the dominant bottlenecks in the weakly thermal phase (small $\delta$) are
those that are locally critical.  Thus, if the typical distance from the critical point is $\delta$, we expect that the density of bottlenecks
is given by $t^{-1/z}$, with $1/z=\eta_c r(\delta,\eta_c)$ giving the Griffiths dynamic exponent $z$;
note that $z$ diverges as the transition is approached.
This density of bottlenecks determines the distance over which information can travel in time $t$.

Thus entanglement typically takes time $\sim l^z$ to spread through the worst bottleneck it encounters in spreading over distance $l$,
and for $z>1$ this dominates the entanglement spreading time.
Energy or particle transport is slower: for example, the charge autocorrelation function or ``return probability''~\cite{agkmd} (which is the inverse of the distance diffused in a time $t$) is given by $\langle n_i(t) n_i(0) \rangle \sim t^{-\beta}$, where $\beta = 1/(z + 1)$. Thus, transport is subdiffusive when $z > 1$.
This subdiffusive transport can be linked to a non-trivial behavior of the a.c. conductivity (via a scale-dependent Einstein relation or a resistor-capacitor model~\cite{agkmd}), which has the low-frequency behavior $\sigma(\omega) \sim \omega^{1 - 2\beta}$, also seen in numerics~\cite{agkmd}. 

\sgch{There is some recent numerical evidence~\cite{lvpgs} that diffusive energy transport coexists with subdiffusive spin transport. We discuss how Griffiths effects can give rise to this coexistence in App.~\ref{mcc}.} 


\section{Griffiths effects in systems with no extensive conserved quantities}\label{floquet}

\begin{figure}[tbp]
\begin{center}
\includegraphics{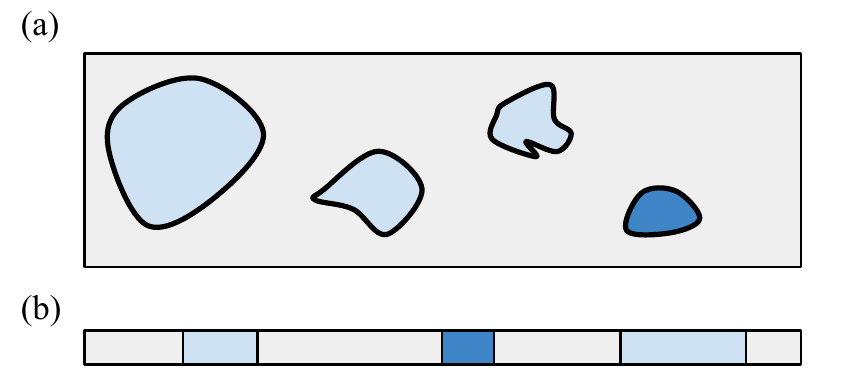}
\caption{Rare region effects in higher dimensions (a) vs. one dimension (b). In all cases, there are insulating inclusions with a wide distribution of sizes and local values of the control parameter (indicated here by shading). In higher dimensions, inclusions can be bypassed, and rare-region contributions are due to degrees of freedom \emph{inside} the inclusions. In one dimension, inclusions act as bottlenecks, and thus affect dynamics even in typical regions.}
\label{dimensionality}
\end{center}
\end{figure}

We now turn from transport to the autocorrelation functions of generic local operators.
%
We shall first discuss these in the conceptually simplest case, which is that of a periodically driven system near a MBL transition, with no extensive conserved quantities (we refer to this as a generic Floquet system).  ``Thermal'' equilibrium for such unconstrained systems maximizes the entropy and thus corresponds in some sense to infinite temperature.
We discuss these in the four separate cases (average vs. typical~\cite{fntyp} and $d = 1$ vs. $d > 1$).

\subsection{Average behavior, any $d$}

The average behavior of generic autocorrelation functions is dominated by rare-region effects.
Starting from \emph{inside} a rare region, the ``escape'' of particles or information from this rare inclusion in to its thermal surroundings will be extremely slow, with timescale $t(L) \sim \exp(L/\eta)$, where $L$ is the shortest linear dimension of the inclusion. More insulating inclusions have smaller $\eta$. \sgch{The rate at which the interior of an insulating inclusion thermalizes with the leads is asymptotically faster than the rate at which the two leads can thermalize ``elastically,'' i.e., without entangling with the inclusion. The matrix element coupling the middle of an inclusion to its edge falls off as $\exp[-L/(2\eta)]$, leading to a Golden-Rule timescale $t(L) \sim \exp(L/\eta)$. By contrast, the matrix element coupling one edge directly to the other will fall off as $\exp(L/\eta)$, which would give rise to a timescale $\sim \exp(2L/\eta)$. The prefactors depend on properties of the leads and are not $L$-dependent, so asymptotically the ``elastic'' process is subleading for insulating inclusions. It is not clear whether this is also true for critical inclusions.} 
Inclusions that are effectively insulating or critical at time $t$ must therefore have a volume of at least $\sim(\eta \log t)^d$, and their density $\sim \exp(-r(\delta,\eta) \eta^d \log^d t)$. Thus (anticipating that these rare regions will dominate the spatial average) we have that

\beq\label{higherD}
[C(t)] \sim \exp(-r(\delta,\eta) \eta^d \log^d t) ~.
\eeq
The inclusions that dominate the long-time behavior are those with local $\eta$ that minimizes $r(\delta,\eta)\eta^d$; this minimum value is the coefficient $\alpha$ in Eq. (1).
The rare-region contribution to all spatially-averaged autocorrelators and dynamical observables will take the form~\eqref{higherD} for operators that do ``freeze'' in the MBL phase.  We note that Eq.~\eqref{higherD} superficially resembles a result from classical spin glasses~\cite{rsp}; however, the physics is different, as we are concerned with the escape from an insulating region and Ref.~\cite{rsp} considers collective domain flips in a spin glass.

\subsection{Typical behavior, $d = 1$}

In one dimension, when the Griffiths dynamic exponent $z>1$ the typical spacing between rare insulating regions is given by $t^{1/z}$,
as noted above, and therefore grows sublinearly in the time $t$ at large $t$.  This gives two related mechanisms by which these rare regions affect
the typical long-time behavior of autocorrelation functions.  The operators within the rare region whose autocorrelations do not decay on
time $t$ will have ``tails'' in the adjacent regions containing typical sites.  Also, on
timescale $t$, any typical part of a system can effectively be regarded as being in a ``box'' of size $L \sim t^{1/z}$ that is isolated (on this timescale) from the rest of the system. Thus, a generic long-time autocorrelation function in such a box will have a value $\agt 1/N(t)$, where $N(t)$ is the Hilbert space dimension of the box---specifically, $N(t) \sim \exp[sL(t)]$ where $e^s$ is the number of states per site.  Thus, in the generic case, the \emph{most} that a typical autocorrelation function can decay on time scale $t$ is given by a ``stretched exponential'':
\beq\label{stretched}
C_{typ}(t) \agt C_{typ}(t = 0) \exp(- \mathrm{const.} \times t^{1/z}) ~.
\eeq
%
Consequently, whenever $z > 1$ (i.e., in the Griffiths regime of Sec.~\ref{review}), the long-time decay of typical autocorrelators is slower than a simple exponential. (We do not rule out the possibility of even slower decay, though generically we expect inequality~\eqref{stretched} to be saturated.) Note that these typical autocorrelations are subleading to \emph{average} autocorrelations, which decay as a power law in $d=1$.

\subsection{Typical behavior, $d > 1$}

For $d>1$, the contributions originating from \emph{inside} rare insulating inclusions do not affect typical behavior as the typical distance to the nearest such inclusion grows superlinearly with $t$. (Thus, on a timescale $t$, a typical site is not within the zone of influence of an inclusion that is insulating on timescales $\sim t$.)  Moreover, because entanglement can spread \emph{around} inclusions in higher-dimensional systems, the inclusions do not act as bottlenecks. Therefore, we can ignore Griffiths phenomena entirely for this case.  Since by assumption there are no hydrodynamic quantities in generic Floquet systems, these typical local autocorrelation functions decay exponentially (but see Ref.~\cite{kim}).

\subsection{Summary and implications for spectral functions}

The discussion above shows that Griffiths effects determine the decay of spatially averaged correlation functions, regardless of dimension,
in systems with no extensive conserved quantities. This is because the average is dominated by correlation functions \emph{inside} the inclusions,
which take a long time to decay. Further, in one dimension, Griffiths effects dominate the decay of \emph{typical} correlations provided
that the density of inclusions is large enough: this is because inclusions act as bottlenecks, inhibiting the equilibration of the
typical regions between them. An important implication of our discussion, specific to the generic Floquet case, is that the coefficient
$\alpha$ (and thus the decay power law) is the \emph{same} for all spatially-averaged local correlators in one dimension when $z>1$,
provided they are correlators of operators that do ``freeze'' in the MBL phase\cite{cases}.
%

We briefly comment on the implications of these results for spectral functions, which we can obtain directly by Fourier transforming the autocorrelation functions discussed above.
When the temporal decay is faster than a power-law (i.e., for averaged correlation functions in higher dimensions, and for typical correlation
functions in one dimension) the spectral functions exhibit at most a weak essential singularity at $\omega = 0$ due to rare regions.
This is on top of the \emph{typical} behavior, which is a smooth function
that grows increasingly sharply peaked at $\omega=0$ as one approaches the MBL transition~\cite{gn} (the width of this reflects the typical relaxation time, which diverges at the transition).  For averaged spectral functions in one dimension, however, the long-time power-law decay implies that the spectral functions have the low-frequency behavior
\beq
[\tilde{C}(\omega)] \sim \mathrm{const.} + \omega^{(1-z)/z},
\eeq
where a constant part due to the typical decay is always present.  Far from the MBL transition, $z < 1$, and this Griffiths power-law is subleading to the constant in spectral functions.  Close to the transition, $z > 1$ and generic local spectral functions exhibit a low-frequency divergence. Note that, as the MBL transition is approached in one dimension, these averaged spectral functions approach the form $\sim 1/\omega$, which is possibly related to recent discussions of $1/f$ noise in disordered spin systems~\cite{adm}.

\section{Griffiths effects in systems with extensive conserved quantities}\label{Hamiltonian}

We now turn to systems with global conservation laws, such as energy or charge conservation. The densities of conserved quantities relax diffusively (i.e., as $\sim t^{-d/2}$ for local autocorrelations) even in generic clean systems; thus there are multiple sources of slow dynamics in these systems. Once again, we address the various cases in turn. \sgch{We focus, in the main text, on the case of a single conserved quantity. In a Hamiltonian system, this must be energy; in a driven system, it can be any quantity conserved by the drive. We discuss the case of multiple conserved quantities in App.~\ref{mcc}; each conserved quantity can in general have a separate value of the localization parameter $\eta$, allowing for the coexistence of normal and anomalous diffusion in one-dimensional systems.}

\subsection{Typical and average behavior, $d > 1$}

In systems with conserved quantities, the rare-region contributions to generic autocorrelation functions continue to take the form~\eqref{higherD}. However, in systems with conserved quantities, these rare-region effects are not the only source of slow dynamics in the system; in addition, there are hydrodynamic modes, corresponding to slow fluctuations of the conserved densities. It is well known~\cite{pr, emdb, lmmr} that these give rise to \emph{long-time tails} in the decay of generic autocorrelation functions; i.e., the \emph{typical} behavior of a generic autocorrelation function is to decay at long time as a power law, which is slower than the rare-region contribution, so that Griffiths effects are subleading in averaged as well as typical autocorrelators in $d > 1$. In fact, only a special set of autocorrelation functions are immune from long-time tails; we discuss how to identify and observe these below.

\subsection{Typical and average behavior, $d = 1$}\label{1dtyp}

In one dimension, both rare regions and hydrodynamics give power-law decay, and---as we now discuss---their effects are intertwined. A generic autocorrelation function contains some overlap with the conserved densities themselves, and these decay as $t^{-\beta} \equiv t^{-1/(z + 1)}$ when $z \geq 1$, as discussed in Sec.~\ref{review}. Autocorrelation functions that do not directly overlap with the conserved densities are nevertheless coupled to these densities~\cite{pr} and thus pick up subleading long-time tails with more rapidly decaying power laws. The \emph{typical} behavior of autocorrelators will generically be sensitive to these subleading long-time tails. On the other hand, the \emph{average} behavior is dominated by the slower of two power-laws: the power-law originating from inside rare regions, and that originating from typical regions. We illustrate these points below by discussing the relaxation of current and density fluctuations as a function of their wavevector $q$.

\subsubsection{Rare-region contribution}

Within an inclusion that is insulating on timescale $t$, generic local operators do not relax at all; thus, their contribution to the spatial average is $\sim t^{-1/z}$, precisely as in the previous section.
(We should specify here that we are considering operators that are even under time-reversal; operators that have the ``wrong'' symmetry,
such as current, decay
inside an inclusion.)

\subsubsection{Local and global optical conductivity}

As a specific case we consider the current-current autocorrelation function, which is related to the optical conductivity by a Kubo formula. For a system that is time-reversal invariant, observables that are odd under time-reversal will generically pick up the long-time behavior of the current, but not the density (which is even under time-reversal). We first consider the behavior of the \emph{local} current, i.e., $\langle j_i(t) j_i(0) \rangle$ at some site $i$. On a timescale $t$, for $z>1$ this site is in effect contained in a box of size $L(t) \sim t^{\beta}$ where $\beta=1/(z+1)$ is the subdiffusion exponent~\cite{agkmd}. Equilibrium density fluctuations imply that typically the density to the left and right of site $i$ differ by $1/\sqrt{L(t)}$. This density imbalance relaxes on a timescale $t$ (which is the timescale for equilibration across $L(t)$), and its relaxation involves moving $\sim \sqrt{L(t)}$ units of the ``charge'' associated with the conserved density across site $i$. Thus the local current-current correlator at site $i$ has the power-law behavior

\beq
\langle j_i(t) j_i(0) \rangle \sim [\sqrt{L(t)}/t]^2 \sim t^{-2 + \beta}
\eeq
Note that, unlike the density-density correlator, this decays \emph{more} rapidly as the MBL transition is approached; this is natural as there are no frozen currents in the MBL phase.

The \emph{total} current \emph{in the region}, denoted $J$, has a slower long-time tail: to relax the initial density imbalance, a net
$\sim \sqrt{L(t)}$ particles must be moved a distance $\sim L(t)$. Including this factor (which can equivalently be seen as multiplying
$j_i$ by the number of sites over which current flow is correlated at time $t$), we get an autocorrelation for the total current of order
$1/t^{2 - 3\beta}$. One can relate this to the a.c. conductivity\cite{kubo}
as follows. Since currents in separate regions of size $L(t)$ are uncorrelated we can just add up the dissipation due to these uncorrelated regions; this amounts to adding up their conductivities~\cite{dyre}. Each \emph{region} has a conductivity that is related to the current-current correlator by

\beq
\sigma(q=0, \omega) \sim \frac{1}{L(1/\omega)} \int dt e^{i\omega t} \langle J(t) J(0) \rangle.
\eeq
This Fourier transform gives the result~\cite{agkmd} that

\beq
\sigma(q, \omega) \sim \omega^{1 - 2 \beta} = \omega^{(z - 1)/(z + 1)} \qquad q L(1/\omega) \ll 1.
\eeq
The above result applies not only to the $q = 0$ conductivity but also to $q > 0$ conductivity provided that $q L(1/\omega) \ll 1$: in this limit the length-scale over which relaxation occurs is governed by $\omega$ rather than $q$.

\subsubsection{Density-wave relaxation, structure factor, and large-$q$ conductivity}

An observable of particular experimental interest \cite{bloch} is the relaxation of a patterned initial state (typically a density
wave of wavenumber $q$).
The measured quantity is the expectation value of this density wave at a later time $t$, denoted $\mathcal{I}_q(t)$.
While this is not a local correlator, it can be analyzed using the same reasoning.  At a time $t$, the density has relaxed over a scale $L(t) \sim t^{\beta}$, but on larger scales the system is cut into segments separated by bottlenecks.  The average deviation from equilibrium of the density in a segment of length $L(t)$ between bottlenecks is  $\sim 1/(q L(t))$, and the corresponding overlap is $1/(q L(t))^2$.  Thus the typical regions contribute an overlap $\sim 1/t^{2\beta}$.  Note that this is always subleading, in the spatial average, to the contributions originating from inside the rare Griffiths regions (because $2 \beta \equiv 2/(z + 1) \geq 1/z$ when $z > 1$). Therefore, the contrast decay goes as $\mathcal{I}_q(t) \sim t^{-1/z}$.

A very similar Griffiths analysis can be performed for the $q$-dependent autocorrelation function of the density, $\hat{S}(q, t) \equiv \mathrm{Tr}[U^\dagger(t) \hat\rho_q U(t) \hat\rho_q \exp(-\beta \hat{H})]$. Between inclusions that are insulating at time $t$, the remaining ``memory'' of the initial density modulation consists of a density excess or deficit of order $1/(qL(t))$ that is spread out uniformly over the scale $L(t)$. Once again, this typical-region contribution to $\hat{S}(q,t)$ goes as $t^{-2\beta}$, and is subleading to the rare-region contribution $\sim t^{-1/z}$ from inside inclusion cores.

Thus the autocorrelator $\hat{S}(q, t) \sim t^{-1/z}$ and its Fourier transform, the structure factor $S(q, \omega) \sim \omega^{1/z - 1}$. Consequently~\cite{igb}, the behavior of the conductivity $\sigma(q, \omega)$ when $q$ is finite and $\omega \rightarrow 0$ goes as

\beq
[\sigma(q, \omega)] \sim \omega^{1 + 1/z} \qquad q L(1/\omega) \gg 1.
\eeq

%

\subsection{Summary}

In this section we argued that, for systems with conserved quantities, hydrodynamic power laws generically mask Griffiths effects,
in both average and typical autocorrelation functions, in $d > 1$.  In $d = 1$, on the other hand, Griffiths power-laws are dominant
sufficiently near the transition.  There are two sources of Griffiths power laws: first, the inclusions themselves directly contribute
(as they do in Floquet systems); second, for $d=1$ the inclusions act as bottlenecks for the transport, which slows the relaxation of
typical regions in between bottlenecks.
Thus, in contrast to Floquet systems, Hamiltonian systems have \emph{different} continuously varying Griffiths exponents for different observables. Moreover, not all exponents vanish near the transition. Indeed, some observables, such as the current, decay \emph{faster} (though still as power laws) near the MBL transition, because they are required by symmetry to vanish in the MBL phase.

In the frequency domain, a generic spectral function will go (when $\omega \rightarrow 0$) as $[\tilde{C}(\omega)] \sim A + B \omega^p$, where $p$ is an exponent related to the temporal long-time tail of the associated autocorrelator. When $p \geq 0$, these power-laws are subleading in the spectral function, though they still dominate the long-time behavior of the autocorrelator.  In contrast with the Floquet case, both typical and average autocorrelation functions have power-law singularities as the transition is approached.  However, the typical and average power laws may differ, with the latter being slower.

The dependence of the conductivity, $\sigma(q, \omega)$, on wavevector $q$ and frequency $\omega$, in one dimension, illustrates many of these features. When $\omega$ is taken to zero keeping $q$ finite, relaxation can occur locally, and the conductivity vanishes with an exponent $\omega^{1 + 1/z}$, due to slow relaxation \emph{within} rare regions. On the other hand, when $q$ is taken to zero keeping $\omega$ finite, relaxation requires large-scale rearrangements of the conserved quantity, and the conductivity vanishes with an exponent $\omega^{(z - 1)/(z + 1)}$, determined by slow relaxation \emph{across} rare regions.

\section{Bypassing long-time tails through spin echo}\label{secse}

\begin{figure}[htbp]
\begin{center}
\includegraphics[width = 0.45\textwidth]{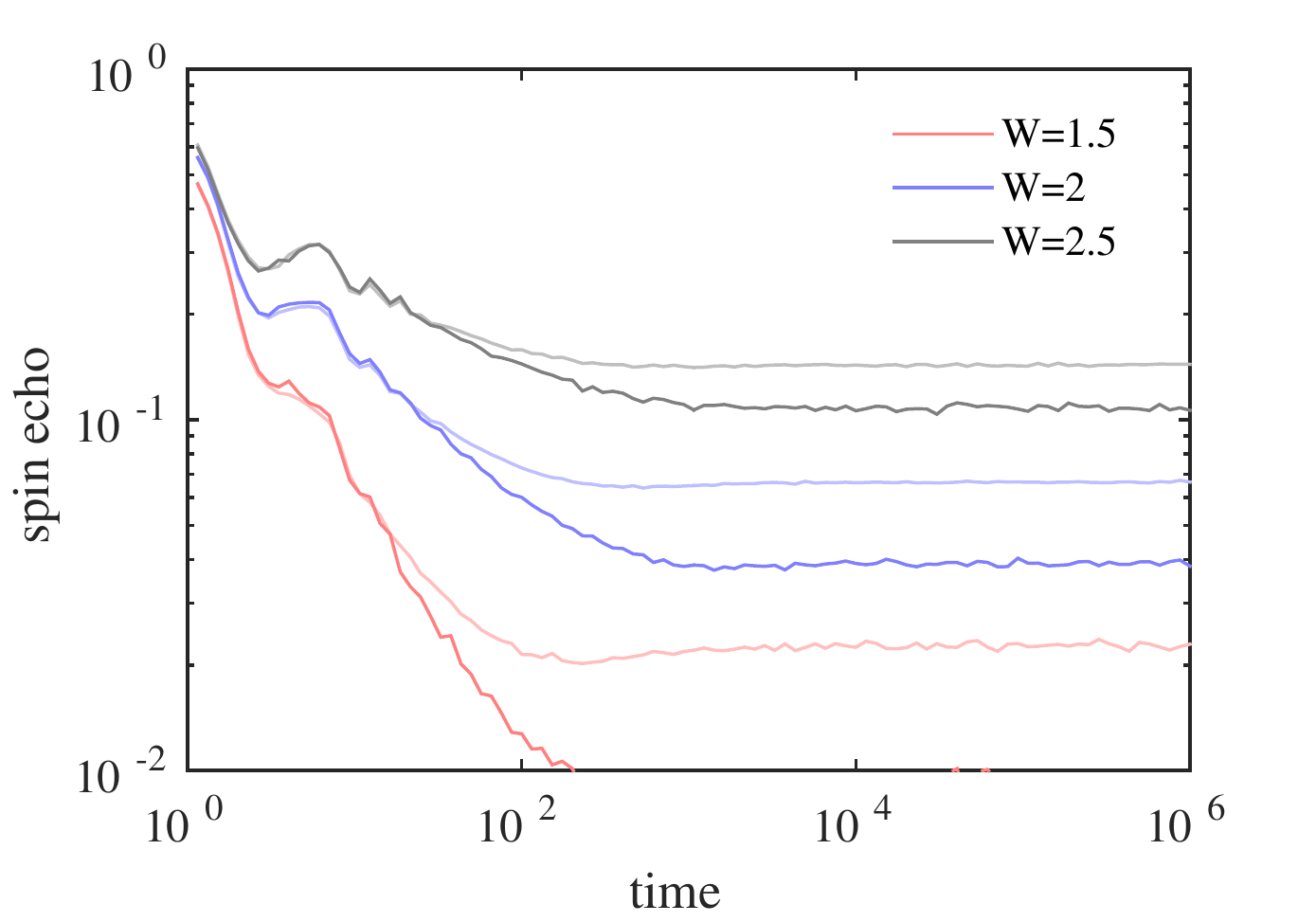}
\caption{Spin echo response for various disorder values $W$ in the thermal phase of the random-field Heisenberg model, $H = \sum\nolimits_i h_i S^z_i + \mathbf{S}_i \mathbf{\cdot S}_{i + 1}$, where $h_i \in [-W, W]$. Thin lines correspond to $L = 12$ (averaged over 10000 realizations) and thick lines to $L = 14$ (averaged over 1000 realizations).}
\label{se}
\end{center}
\end{figure}

Although \emph{generic} autocorrelators exhibit long-time tails for systems with conserved densities, in some cases it is possible to construct simple operators that do not.  A specific class of such quantities are ``transverse'' operators $O_\perp$ that change the value of a discrete conserved quantity, such as single-particle creation operators [or, e.g., in XXZ spin models, spin projections that are perpendicular to the conserved one].  Using the method of fluctuating hydrodynamics~\cite{pr}, one can argue that autocorrelation functions of the form $\langle O_\perp^\dagger(t) O_\perp(0) \rangle$ decay exponentially even after nonlinear hydrodynamic effects are included. \sgch{ The argument is as follows: long-time tails in the autocorrelations of an operator $O$ arise because of mixing between that operator and the slow operators of the theory, i.e., conserved densities (or their products, derivatives, etc.). If we denote some particular slow operator by $Q(t)$, then the extent of mixing between $O$ and $Q(t)$ is governed by the operator inner product~\cite{pr, mori}}

\beq
( O | Q(t) ) \propto \mathrm{Tr}[O Q(t)]
\eeq
where for convenience we have chosen time labels such that $O \equiv O(t = 0)$. The slow operators $Q(t)$ act only within a particular sector of the global conserved quantity, whereas the transverse operator $O_\perp$ by definition changes the value of the global conserved quantity. Thus, $(O_\perp | Q) = 0$ for a transverse operator, and consequently these purely ``transverse" autocorrelators do not pick up long-time tails.  Unfortunately such operators are also orthogonal to the emergent conserved quantities in the MBL phase; consequently, their autocorrelation functions in the MBL phase will precess at a state-dependent frequency, and thus decay upon spatial averaging~\cite{DEER}.

%

This decay of the autocorrelations within the MBL phase can be undone using spin echo~\cite{DEER}; we now argue that spin echo in the thermal Griffiths regime is dominated by rare-region contributions in all dimensions. For specificity, we consider a system of spins-1/2 with a global $U(1)$ symmetry, corresponding to a conserved spin projection,
which we label $z$. The generalization to bosonic and fermionic systems with particle number conservation is straightforward (see Appendix A).
In general, a local spin-flip operator $\sigma_i^x$ in such a system in the MBL phase will have nonzero overlap
with one or more operators $\tau^x_j$ that flip a single conserved pseudospin.

In the spin-1/2 case with a globally conserved $z$ magnetization, the spin echo response (or ``fidelity'') at a site $i$ can be written~\cite{DEER} as
\bea
\mathcal{F}(t) & = & \langle \psi(t) | \sigma^z_i | \psi(t) \rangle ~, \nonumber \\
| \psi(t) \rangle & = & -\frac{1}{4}(1 - i \sigma^y_i) e^{-i H t/2} (1 - i \sigma^y_i)^2 \\ && \qquad \times e^{-i H t/2} (1 - i \sigma^y_i) |\psi(0) \rangle ~.\nonumber
\eea
The spin echo response is closely related to the autocorrelation function of the non-conserved components of the spin.  As such, $\mathcal{F}(t)$ decays to zero at long time
in the thermal phase, while it saturates to a finite value 
in the localized phase. This is the same as the behavior of the generic autocorrelator we discussed above; thus, we once again arrive at the rare-region contribution~\eqref{higherD} to the spin echo response. This power-law decay of the response indeed seems to arise in the random-field Heisenberg chain (Fig.~\ref{se}), though, as is typical in the thermal phase, our numerical results have strong finite-size effects that arrest the decay after some finite, $L$-dependent time.

\section{Nature of dominant rare regions}\label{dominantregions}

To establish the functional form of the rare-region contribution, we did not need to address the question of what the dominant inclusions are \emph{like}: i.e., whether they are locally critical or insulating, and by how much.  However, the nature of these inclusions determines the factor $\alpha$ in the exponent in Eq.~(1); in one dimension, this sets the power-law with which correlation functions decay.

To address this question in some generality, we will consider various possible scalings of the time for information or particles to
escape a critical inclusion of size $L$ embedded in a thermal bulk background.
For an \emph{insulating} inclusion, this time goes as $t(L)\sim\exp(L/\eta)$, where $L$ is the shortest dimension of the inclusion.
Since a critical inclusion must relax \emph{faster} than an insulating inclusion, the possibilities for critical dynamics are
(i) that $t(L)$ remains exponential in the length (i.e., $t(L) \sim \exp(L/\eta_c)$), or (ii) that it grows sub-exponentially in
$L$ (e.g., as a power-law of $L$, or as $\exp(\kappa L^\psi)$ with $\psi < 1$).  In one dimension, as discussed in Sec.~\ref{review}, there is evidence
from both numerical and renormalization-group methods that possibility~(i) obtains.  In higher dimensions, there is no direct evidence
either way, although possibility~(i) seems more plausible~\cite{fninc}.

In what follows, we discuss in general terms how these assumptions determine the behavior of the Griffiths prefactor/exponent $\alpha$ in Eq.~\eqref{eq1} (in Appendix~\ref{appB} we specialize to the one-dimensional case, and discuss the leading corrections to the behavior we have seen). We shall take the general form $\exp(\kappa L^\psi)$ for critical dynamics ($0 \leq \psi \leq 1$), which includes all the cases of interest. The relaxation rate of a localized inclusion will be $t(L, \hat{\xi}) \sim \exp(L/\eta(\hat{\xi}))$ when $\hat{\xi} \leq L$, and $t(L) \sim \exp(\kappa L^\psi)$ for critical inclusions where $L \leq \hat{\xi}$, where $\hat{\xi}$ is the local correlation length within the inclusion.  Matching these regimes gives us that $\eta(\hat{\xi}) \sim \hat{\xi}^{1 - \psi} \sim |\hat{\delta}|^{-\nu(1 - \psi)}$. Now, suppose the system is in the thermal phase and typically at a distance $\delta >0$ from the critical point.  A localized inclusion with internal control parameter $\hat{\delta} <0$ gives a contribution

\beq
\exp(- \eta(\hat{\delta})^d r[\delta, \eta(\hat{\delta})] \log^d t)
\eeq
to the autocorrelation. To find the dominant inclusions, we therefore need to minimize the quantity $\eta^d r(\delta, \eta)$. We now use the critical behavior $\eta \sim |\hat{\delta}|^{-\nu(1 - \psi)}$, the fact that $\hat{\delta}$ is itself a local property, and the small-argument behavior of the rate function from Sec.~\ref{notation} to find that

\beq
\eta^d r(\delta, \eta) \sim \delta^\rho (\delta - \hat{\delta})^2 |\hat{\delta}|^{-\nu d (1 - \psi)}
\eeq
where $\rho$ is the exponent defined in Sec.~\ref{notation}, which satisfies $\rho > -1$.
Let us take $\delta$ to be in the thermal phase, a small distance from the critical point, and find the dominant $\hat{\delta}$.
Two kinds of behavior are possible, depending on the value of $\nu$. In particular, we see that

\beq\label{griffiths_req}
\nu d (1 - \psi) < 2 \Rightarrow \alpha \xrightarrow{\delta \rightarrow 0} 0.
\eeq
In this case, the dominant $\hat{\delta}$ is near-critical when $\delta$ itself is near-critical.
On the other hand, if $\nu d (1 - \psi) > 2$, the optimal inclusions remain deeply insulating all the way to the critical point
(although $\alpha$ can still vanish if $\rho > 0$).
The cases $\psi = 1$ (corresponding to $t(L) \sim \exp(L/\eta_c)$, which seems most likely to be true) and $\psi = 0$ (corresponding, e.g., to a finite dynamical critical exponent $z$) are special. When $\psi = 1$, inequality~\eqref{griffiths_req} is always satisfied and the dominant inclusions are always near-critical. When $\psi = 0$, the inequality is always violated (because $\nu d \geq 2$ in disordered systems~\cite{ccfs, clo}) and the dominant inclusions are deeply insulating.

\section{Conclusions}\label{summary}

In this work we have extended previous results on Griffiths effects on the thermal side of the MBL transition from one dimension to higher
dimensions and from transport to spin echo and other dynamical observables.
%
We have identified various considerations that determine whether a given observable and/or system will exhibit a thermal Griffiths regime
where the long-time behavior is dominated by rare regions.
To summarize, our main conclusions are:

(a)~In systems with no conserved quantities, the long-time behavior of thermally and spatially averaged autocorrelators takes the form~\eqref{eq1}
and is dominated by Griffiths effects that are due to slow relaxation \emph{inside} rare regions that are locally insulating or critical.
The coefficient $\alpha$ in Eq.~\eqref{eq1}
is the same for all autocorrelators.  Autocorrelators at typical spatial locations decay parametrically faster (exponentially in $d > 1$ and with stretched exponentials due to insulating bottlenecks in $d = 1$).

(b)~In systems with conserved quantities, when $d > 1$, the long-time behavior of generic autocorrelators is dominated by hydrodynamic tails. The Griffiths behavior~\eqref{eq1} can be recovered either as an intermediate-time transient, or by choosing \emph{specific} measurements, such as spin echo, for which hydrodynamic long-time tails are absent. When $d = 1$, Griffiths effects dominate general autocorrelators near the transition, but the Griffiths exponents are modified by hydrodynamic effects.

Although our discussion has focused on MBL systems with short-range interactions, it can directly be extended to systems with longer-range (e.g., power-law~\cite{burin-old, yao2013, burin2015} or stretched exponential~\cite{va, vpp}) interactions, provided that the interactions fall off fast enough for an MBL phase to exist. Such systems avoid a subdiffusive phase in all dimensions, and Griffiths effects in them are qualitatively similar to those in short-range systems with $d > 1$.
An interesting question is to what extent the Griffiths effects discussed here extend to systems with correlated disorder. To give an extreme instance, many-body localization can occur in systems without quenched randomness that are subject to quasiperiodic potentials \cite{iyer,bloch}. Within the thermal phase, we do not expect Griffiths effects of the type discussed here to play a significant role in this limit of highly correlated disorder; however, the fate of the subdiffusive phase as the disorder correlations are made long-range is currently unclear. 

These results for the thermal phase, with their strong dependence on dimensionality and the existence of conserved quantities, contrast markedly with Griffiths effects within the MBL phase. Throughout the MBL phase, response is dominated by locally atypical regions, either of the disorder configuration or of the state (thus, again, quasiperiodic and random systems can be understood on the same footing). However, the rare-region effects in the MBL phase appear to be dimension-independent, and always give rise to power laws in the dynamics~\cite{mbmott}.  Thus, a MBL transition in higher dimensional systems would have the intriguing feature that rare region effects are dominant throughout the localized phase, but subleading throughout the thermal phase. The implications of this for the critical behavior at the phase transition will be addressed in future work.

\emph{Note added}.---As this manuscript was being prepared, a numerical study appeared~\cite{lfa} providing evidence for anomalous Griffiths effects in the imbalance decay (cf. Sec.~\ref{1dtyp}).

\section{Acknowledgments}

We thank E. Altman, F. Huveneers, M. M\"uller, A. Potter, U. Schneider, and especially V. Oganesyan for helpful discussions.
The authors acknowledge support from Harvard-MIT CUA, NSF Grant No. DMR-1308435, AFOSR Quantum Simulation MURI, the ARO-MURI on Atomtronics, ARO MURI Quism program.
S.G. acknowledges support from the Walter Burke Institute at Caltech and from the National Science Foundation under Grant No. NSF PHY11-25915. D.A.H. is the Addie and Harold Broitman Member at I.A.S.  M.K. acknowledges support from Technical University of Munich - Institute for Advanced Study, funded by the German Excellence Initiative and the European Union FP7 under grant agreement 291763.
E.D. acknowledges support from the Humboldt Foundation, Dr.~Max R\"ossler, the Walter Haefner Foundation and the ETH Foundation.

\appendix

\section{Coexistence of normal and anomalous diffusion} \label{mcc}

In this Appendix we discuss transport in one-dimensional systems with multiple conserved quantities near the MBL transition. For simplicity, we consider a toy model consisting of a system with two types of excitations, ``neutral'' (i.e., carrying energy but no charge) and ``charged'' (i.e., carrying energy and charge). 
We take the interactions between these two types of excitations to be weak compared with the characteristic local bandwidth of either excitation. 
With these assumptions it is clear that the only possibilities are for both types of excitation to be localized or for both to be delocalized: delocalization in one sector spreads to the other in the presence of interactions, because each sector acts as a ``bath'' for the other~\cite{hyatt, pg}. Thus, it seems that a diverging localization length in one sector must imply the same for the other. Nevertheless, the numerical values of the localization length (and thus of the parameters $\eta$ defined in the main text) need not in general be the same for both types of excitation. 

Let us first consider a limit in which the two types of excitation are entirely decoupled. Then in general, an inclusion of size $L$ that is insulating or critical for both neutral and charged excitations and is embedded in a thermal background will have separate transit times $t_n(L) \sim \exp(L/\eta_n)$ for neutral excitations and $t_q(L) \sim \exp(L/\eta_q)$ for charged excitations. Thus when $\eta_q \ll \eta_n$, this model can have subdiffusion of charge together with diffusion of energy, which is the situation seen numerically in Ref.~\cite{lvpgs}. Note that the ratio $t_q(L)/t_n(L) \sim \exp[L (\eta_q^{-1} - \eta_n^{-1})]$, which grows exponentially with the size of the inclusion. 

We now investigate the stability of this situation when the two types of excitations are weakly coupled. In addition to the direct process (involving the transmission of charged excitations through the inclusion), it is now also possible to have ``hopping transport,'' i.e., real transitions in the charge sector that borrow energy from the ``bath'' provided by the more rapidly relaxing neutral sector. In the middle of the inclusion, the effective bath due to the neutral sector has a correlation time $t_n(L) \sim \exp(L/\eta_n)$. Thus, when $L$ is large, the local bath is ``slowly fluctuating'' in the sense of Ref.~\cite{gn}. We can then use the results of that work to conclude that the charge rearrangement rate 

\beq
\Gamma^{hop.}_q \simeq t(L)^{1 - 2/(s\zeta_q)} \simeq \exp\left[-L \left( \frac{2/\phi_q - 1}{\eta_n}\right)\right].
\eeq
The expression $\phi_q$ is the coefficient of the exponential phase-space growth in the MBL phase~\cite{mbmott, gn}, with the properties that $\phi_q \rightarrow 0$ deep in the MBL phase and $\phi_q \alt 1$ everywhere inside the MBL phase. (The parameter $\phi_q$ is conceptually distinct from $\eta_q$, being a dimensionless scale rather than a length.)
Thus, both the direct and the ``hopping'' channel give rise to charge transport that is parametrically slower (specifically, exponentially slower in $L$) than energy transport, provided that $\phi_q \ll 1$, $\eta_q \ll \eta_n$ (i.e., when the charge excitations in isolation would be well localized).

This reasoning can be extended to the case of systems that have local charge hopping but power-law density-density interactions~\cite{yao2013}, which are assumed to be sufficiently rapidly decaying that MBL persists. Naively, one might think that as the charge hopping is short-range, charge transport through an insulating inclusion of length $L$ should be exponentially slow in $L$. In such situations, the slowly fluctuating bath of energy excitations provides a parametrically faster relaxation channel. Following the logic of the previous paragraph, $t(L) \sim L^a$, where $a$ is the power law. Thus, 

\beq
\Gamma_q^{(PL)} \simeq L^{a(2/(s\zeta_q) - 1)}
\eeq
Thus, subdiffusion does not occur in systems with local charge motion but long-range density-density interactions. 

\section{``Spin echo'' for bosonic and fermionic systems}\label{appA}

In bosonic or fermionic systems that have a conserved particle number, the main apparent obstacle to implementing spin echo is that the natural analog of a $\pi/2$ pulse involves creating superpositions of states with different total particle number. In cold-atom experiments, such superpositions \emph{can} straightforwardly be created, as discussed, for example, in Refs.~\cite{ksn, kkg, ksg}. The essential idea is to trap two different hyperfine states of the atoms with a strongly state-selective potential: for instance an experiment might involve hyperfine state $a$, which is used to realize the many-body physics of interest, and a ``spectator'' hyperfine state $b$, which contains very few atoms. The potential experienced by the atoms in state $b$ is strong enough to confine them to a single site or a few sites. Given this setup, driving radio-frequency pulses of the appropriate duration between states $a$ and $b$ can be used to create local superpositions with different ``particle number'' (i.e., different numbers of $a$ particles). The rest of the spin echo sequence can be implemented as usual, and can be checked to saturate to a finite value deep in the MBL phase.
Note, however, that this saturation value need not be near unity, especially for softcore bosons, because a $2\pi$ pulse does not correspond to the identity (but might also involve injecting or removing two particles from the system).

\section{Leading finite-time corrections in one dimension} \label{appB}

In this Appendix we discuss the leading corrections to the long-time asymptotic behavior analyzed in the main text. We argue that these corrections can lead to systematic overestimates of the Griffiths dynamical exponent $z$.  In particular, entanglement (energy) spreading at long but finite times might seem sub-ballistic (sub-diffusive) even when the asymptotic behavior is ballistic (diffusive).  These corrections might account for the surprisingly large size of the anomalous ($z > 1$) regime seen in finite-time numerical studies~\cite{blcr,lfa}.  For concreteness and to make contact with numerics, we focus on one-dimensional systems and make the assumption (motivated by numerical~\cite{ph} and renormalization-group studies~\cite{VHA, ppv}) that the relaxation time for a critical inclusion of size $L$ is given by $t(L) \sim \exp(L/\eta_c)$.

We consider two sources of finite-time corrections: (i)~subleading contributions to the finite-time averages of various observables, and (ii)~corrections that arise because the optimal internal control parameter, $\hat{\delta}$, for a rare region is itself a function of the size of that region, and therefore implicitly of time.

\subsection{Corrections due to averaging}

The conceptually simpler of these issues can be understood as follows. Let us consider the growth of entanglement across a particular cut in the system, starting from a product state~\cite{lfa}. Specifically, we imagine averaging the (von Neumann) entanglement entropy at time $t$ over cuts and/or disorder realizations, and denote this averaged quantity $[S(t)]$. At short times, the system explores only the immediate vicinity of the cut, so that $[S(t)] \simeq [v] t$, where $v$ is a local ``Lieb-Robinson speed'' for entanglement spread in the vicinity of the cut.  Note that $[v]$ can be interpreted equivalently as a disorder-average or a spatial average.  This average is not dominated by the bottlenecks due to Griffiths inclusions, since they simply have a very small local $v$.  By contrast, at long times, entanglement in a given sample has spread through many regions with different local speeds, and its spread can be limited by the slowest regions it encounters.  Thus the typical \emph{single-sample} value of $S(t) \sim [1/v]^{-1} t$, and can be dominated by bottlenecks where the local $1/v$ is extremely large.  (This is analogous to the standard observation that conductances add at high frequencies whereas \emph{resistances} add at low frequencies~\cite{agkmd}.)  Note that $[1/v]^{-1} \leq [v]$, so the slope of the $[S(t)]$ vs. $t$ curve will necessarily decrease with time.  This crossover from a large slope at short times to a smaller slope at long times will give an apparent exponent smaller than unity even when the true long-time velocity $[1/v]^{-1}$ is nonzero.  More generally, we expect that it can lead to systematic overestimates of the exponent $z$ (underestimates of $1/z$) in numerics.

We now discuss this crossover in more detail, focusing on the case where the Griffiths dynamic exponent satisfies $z < 1$, so the bottlenecks are subleading to simple ``ballistic'' entanglement spread and $[1/v]$ remains finite.  In a single sample the average speed of spread out to time $t$ is given by spatially averaging inverse velocities \emph{over the distance entanglement has spread} in that sample: we denote this as $v_t \equiv \langle 1/v \rangle^{-1}_{t}$.  The disorder average is then the arithmetic average over disorder realizations of $v_t$, we denote this $[v_t]$. (This prescription clearly reproduces the limiting cases above.)  We are interested in how $[v_t]$ approaches its (here, nonzero) infinite-time limit, $v_\infty \equiv [1/v]^{-1}$.

It is helpful to work with the probability distribution, $P(1/v)$, which has a long tail $\sim (1/v)^{-1-1/z}$ due to the rare bottlenecks.  At a late but finite time $t$, this distribution is effectively cut off at $1/v \sim t$, because slower bottlenecks cannot be resolved at this time.  Thus,

\beq
[1/v] - \langle 1/v \rangle_t \simeq \int_t^\infty \frac{1}{v}\frac{1}{v^{-1 - 1/z}} d(1/v) \sim t^{1 - 1/z}~.
\eeq
Consequently, the average speed up to time $t$, $[v_t]$, also converges to its asymptotic value with a finite-time correction that vanishes at long time as $\sim t^{1- 1/z}$.  For $z$ near to, but just below, one this gives a strong and slowly decreasing finite-time correction, which
can give rise to an apparent entanglement growth that appears sub-ballistic even when the asymptotic long-time behavior is ballistic.

Note that these crossovers are specific to the physical quantity that is being averaged: the artifacts discussed here would not arise if we were looking at a quantity such as contrast decay or spin echo, for which the typical-region contribution decays rapidly rather than growing rapidly at short times. Thus, this effect could cause apparent violations of scaling relations between exponents in numerical studies.

\subsection{Corrections due to size-dependence of optimal inclusion type}

\begin{figure}[htbp]
\begin{center}
\includegraphics{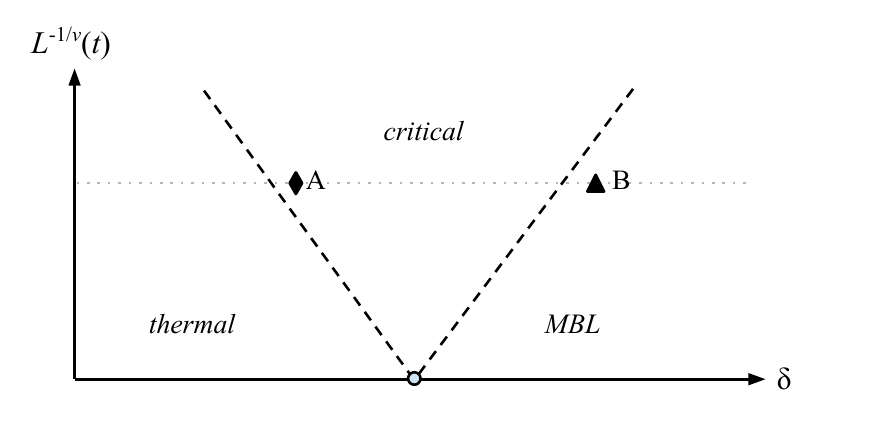}
\caption{Nature of dominant inclusions, assuming $\psi = 1$ critical dynamics (see main text for definition).  The system is globally in the thermal phase; for a given $L$, type-A inclusions are the critical inclusions with highest probability, whereas type-B inclusions are the localized inclusions with highest probability.}
\label{inclusiontypes}
\end{center}
\end{figure}

For $\psi = 1$ critical dynamics, the dominant inclusions that govern dynamical observables are asymptotically \emph{critical}, in the sense that their local control parameter $\hat{\delta} \rightarrow 0$ as their size $L \rightarrow \infty$, even when the global control parameter $\delta$ is slightly in the thermal phase.  At finite $L$, one must distinguish between two types of asymptotically critical inclusions (see Fig.~\ref{inclusiontypes}): (A) inclusions that are internally critical, so that $L \ll \xi(\hat{\delta})$, and (B) inclusions that are internally slightly in the localized phase, so that $\xi(\hat{\delta}) \ll L$, but $\xi(\hat{\delta}) \rightarrow \infty$ as $L \rightarrow \infty$.  Type-A inclusions are the dominant bottlenecks for entanglement and energy spread, as well as for autocorrelation functions whose saturated value in the MBL phase is a power law of $\xi$ or larger. Type-B inclusions dominate the behavior of autocorrelation functions that saturate, in the MBL phase, at values that are exponentially small in $\xi$. Although the rate functions for the densities of type (A) and type (B) inclusions asymptotically approach the same value, the finite-time corrections are different in the two cases, and are slow functions of $\log{t}$, as we now discuss.  The key idea is as follows: an inclusion of size $L$ with internal localization length $\hat{\xi} \agt L$ is effectively critical.  Thus the highest probability type (A) critical inclusions are those with $\hat{\delta}$ slightly thermal and $L(t) \simeq \hat{\xi}$, so that their local control parameter $|\hat{\delta}(L)| \sim L(t)^{-1/\nu}$.  These
are \emph{more} probable than an inclusion with strictly critical control parameter.  The probability of a type (A) critical inclusion of size $L$ is thus given by $\sim \exp[- (r_c - \kappa L^{-1/\nu}) L]$, with $r_c>0$ and $\kappa>0$.  Since $L \sim \log t$, the finite-time spread, for example, of entanglement bottlenecked by type-A critical inclusions will be of the form
\beq
S(t) \sim t^{(1/z) - b (\log t)^{-1/\nu}}~,
\eeq
with $b>0$, so finite-time studies will in general see an apparent power-law spread that is slower than the true asymptotic power law, with the correction vanishing with time only as this small power of $\log{t}$.

If there are Griffiths effects that are instead dominated by type (B) inclusions, then at finite time these inclusions are more rare than critical inclusions, so the finite-time results will in this case give an {\it underestimate} of the asymptotic Griffiths exponent $z$, with the finite-time correction again vanishing only as a slow power of $\log{t}$.





\begin{thebibliography}{99}
\bibitem{fntyp}
For the effects discussed here, one can interpret ``typical'' either in its colloquial sense or the more technical sense of geometrically averaged (i.e., log-averaged) \dhch{or median} quantities.
\bibitem{cases}
There are some special cases where correlations do not have Griffiths effects in Floquet systems, due to the operator having expectation value zero in all eigenstates. One example is purely imaginary Hermitian operators in systems where the eigenstates can all be chosen to be real---an example is the operator $\sigma^y$ in the transverse-field Ising model $H = \sum\nolimits_i h_i \sigma^x_i + \sum\nolimits_{\langle ij \rangle} J_{ij} \sigma^z_i \sigma^z_j$. In the MBL phase such operators precess with a state-dependent frequency rather than freezing. This precession in locally MBL regions averages out and does not give rise to Griffiths effects.
\bibitem{fninc}A degree of freedom in a critical inclusion of size $L^d$ can relax either by decaying directly into the bath (on a timescale exponential in $L$) or by participating in a collective rearrangement involving $L^\theta$ spins (with $0 < \theta \leq d$), on a timescale that is exponential in $L^\theta$. If $\theta > 1$, then the dominant relaxation process is direct decay into the bath. When $d$ is sufficiently large, one expects $\theta > 1$ because the multiplicity of such clusters is larger.
\end{thebibliography}
\end{document}